\documentclass[preprint,number,sort&compress,12pt]{elsarticle}
\usepackage{epsfig}

\usepackage{amssymb}
\usepackage{amsmath}
\usepackage{caption}
\usepackage{subcaption}
\usepackage[per-mode=symbol]{siunitx}

\usepackage[colorlinks=true,allcolors=blue]{hyperref}

\makeatletter
\providecommand{\doi}[1]{%
  \begingroup
    \let\bibinfo\@secondoftwo
    \urlstyle{rm}%
    \href{http://dx.doi.org/#1}{%
      doi:\discretionary{}{}{}%
      \nolinkurl{#1}%
    }%
  \endgroup
}
\makeatother

\newcommand{\refeq}[1]{Eq.~(\ref{#1})}

\newcommand{\reffig}[1]{Fig.~\ref{#1}}

\newcommand{\refsec}[1]{Section~\ref{#1}}

\newcommand{\reftab}[1]{Table~\ref{#1}}
\newcommand{\refref}[1]{Ref.~\cite{#1}}

\begin{document}

\title{Profiles of energetic muons in the atmosphere}
\author{Thomas K. Gaisser}
\address{Bartol Research Institute and Dept. of Physics and Astronomy\\
University of Delaware, Newark, DE, USA}
\ead{gaisser@udel.edu}
\author{Stef Verpoest}
\address{Dept. of Physics and Astronomy, University of Gent, B-9000 Gent, Belgium}
\ead{stef.verpoest@ugent.be}
\begin{abstract}
The production spectrum of high-energy muons as a function of depth in the atmosphere is 
relevant for understanding properties of event rates in deep detectors.
For a given atmospheric profile, cascades of heavy nuclei develop at higher altitude than
proton showers, giving rise to larger separation of muons at depth.  For a given type
of primary cosmic ray, seasonal variations in muon rates reflect the fact that
higher temperatures correspond to lower densities and to a relative increase in the
ratio of decay to re-interaction of the parent mesons.  In this paper, we present a 
generalization of the Elbert formula that tracks meson decay to muons along the 
trajectory of the primary cosmic-ray nucleus.  The convolution of the production 
spectrum with a changing atmospheric profile provides the dependence of 
 event rates and sizes of muon bundles on temperature and primary mass.  We consider applications
 to IceCube and also to multiple muon events in the compact underground detectors of
 MINOS and the NOvA Near Detector.
\end{abstract}

\maketitle

\section{Introduction}

The standard approach to seasonal variations uses formulas for the inclusive 
production spectrum of muons 
integrated over the primary spectrum~\cite{Barrett:1952abc}.  The effective temperature 
$T_{\rm eff}$ is obtained by weighting the production spectrum of muons as a function of atmospheric 
depth with the temperature profile.  The observed variation in the rate $I_\mu$ is then related to the
variation in temperature by a correlation coefficient 
$\alpha_T$ according to
\begin{equation}
\frac{\delta I_\mu}{\langle I_{\mu}\rangle} = \alpha_T\times\frac{\delta T_{\rm eff}}{\langle T_{\rm eff}\rangle},
\label{eq:alphaT}
\end{equation}
where the averages are typically taken for a full year.  
A general expression for the rate is 
\begin{equation}
I_\mu(E_{\mu,\mathrm{min}},\theta)=\int{\rm d}X\int_{E_\mu,\mathrm{min}(\theta)}A_{\mathrm{eff}}(E_\mu,\theta)P(E_\mu,\theta,X){\rm d}E_\mu,
\label{eq:rate0}
\end{equation}
where $E_{\mu,\mathrm{min}}(\theta)$ is the minimum energy for a muon to reach the detector from zenith angle $\theta$.~\footnote{The detectors we consider in this paper all have flat overburdens.}
Here $X$ (\si{\gram \per \cm \squared}) is slant depth from the top of the atmosphere along a trajectory at zenith angle $\theta$, $A_{\mathrm{eff}}$ is the projected effective area of the detector, and $P_\mu$ is the muon production spectrum.
The full rate is obtained by integrating \refeq{eq:rate0} over zenith angle.

In a large
volume detector like IceCube~\cite{Desiati:2011hea,Tilav:2019xmf}, the depth dependence of 
the response is accounted
for with an energy-dependent effective area, and the production spectrum differential
in energy must be used.  For compact detectors like 
MINOS~\cite{Adamson:2009zf}, however, the effective area is the physical area of the detector projected in the direction $\theta$ and averaged over azimuth. It depends only on $E_{\mu,\mathrm{min}}(\theta)$ and factors out of the integral in \refeq{eq:rate0}.
In both cases, the formulas are ``inclusive"
in the sense that the primary spectrum has been integrated over to obtain the flux of
muons per \si{\m \squared \s \steradian}. 

In reality, underground detectors measure rates of events.  This is especially the
case for a large detector like IceCube, where high-energy events consist of large
muon bundles from cores of air showers~\cite{IceCube:2019hmk}.  It is also the case for a tracking 
detector like MINOS when events are characterized by muon multiplicity 
and separation~\cite{Adamson:2015qua}.  The goal of this paper
is therefore to characterize the rate of events in terms of primary cosmic-ray energy
and mass, while accounting for energy, multiplicity and height of origin 
of the muons.     

The paper has three main sections.  
The first describes how muon production can be
parameterized as a function of atmospheric depth and how the formulas integrated
over depth relate to the Elbert formula~\cite{Elbert:1979abc,Elbert:1979gz}.
The following section considers
coincident events in which a surface array provides an indication of the primary particle
while the deep detector measures the properties of the muon bundle.
The final section deals with underground measurements in which the weighted sum of
all primaries is taken.  The emphasis of this section is on seasonal variations
of events in underground detectors and their dependence on muon multiplicity.

\section{Simulations and fitting parameters}
\label{sec:parameters}

\begin{figure}
\includegraphics[width=0.6\textwidth]{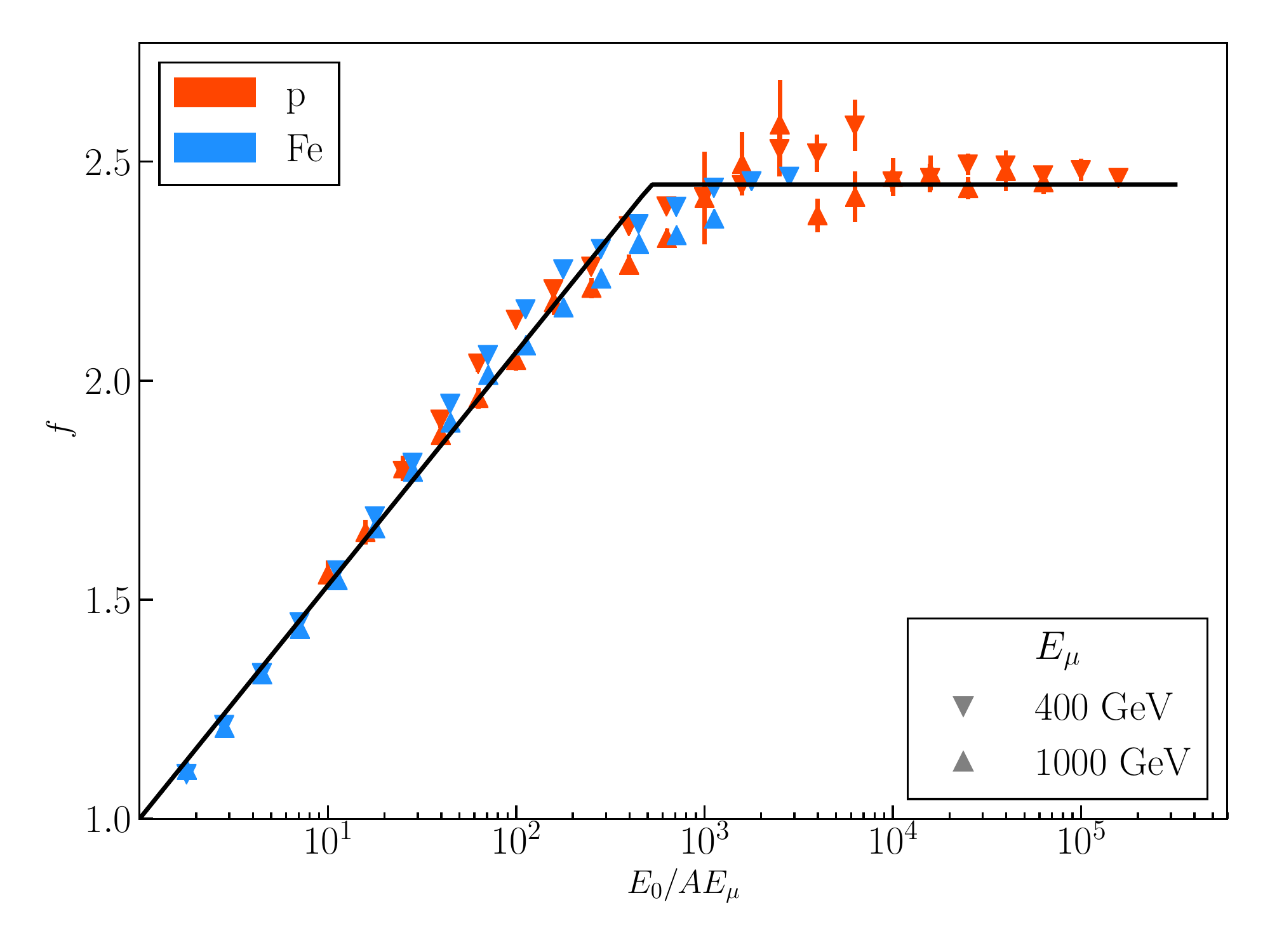}
\centering
\caption{Value of the ratio $f$ of the minimum and mean (above minimum) muon energy in a shower. The black line is an approximate description of the values inferred from simulations of vertical proton and iron primary cosmic rays over a large range of primary energies and with two different minimum muon energies.}
\label{fig:factor}
\end{figure}

The formula originally proposed by Elbert~\cite{Elbert:1979abc,Elbert:1979gz} as an approximation to the number of
high-energy muons produced per primary cosmic ray, has been used to estimate properties
of muon bundles in deep underground detectors~\cite{Gaisser:1985yw,Forti:1990st}.  
A standard form~\cite{Gaisser:2016uoy} is
\begin{equation}
\langle N_\mu(>E_\mu, E_0, A, \theta)\rangle\;\approx\;A\times\frac{K}{ E_\mu\,\cos\theta}\,\left (
\frac{E_0}{ A\,E_\mu}\right )^{\alpha_1}\,
\left (1\,-\,\frac{A\,E_\mu}{E_0}\right)^{\alpha_2},
\label{eq:ElbertFormula}
\end{equation}
where $A$ is the mass number of a primary nucleus of total energy $E_0$, and the values of the normalization constant $K$ and exponents $\alpha_1$ and $\alpha_2$ are included in the tables of parameters below.
The scaling with $A\,E_\mu/E_0$ follows from the superposition approximation, in which
incident nuclei are treated as $A$ independent nucleons each of energy $E_0/A$.
In this paper we generalize the Elbert formula to obtain the distribution of slant depths over
which the muons are produced.  The integral of this distribution is the mean number of
muons per shower, to be compared with \refeq{eq:ElbertFormula}.
The basic idea is to interpret the derivative of the Gaisser-Hillas (G-H) function~\cite{GaisserHillasabc} 
as the rate of production of charged mesons per d$X$~(g/cm$^2$) along the trajectory
of a primary cosmic ray and then multiply by appropriate factors to get the production
spectrum for muons with energy $>E_\mu$,
\begin{eqnarray} 
\label{eq:formula}
\frac{{\rm d}N}{{\rm d}X}(>E_\mu,E_0,A,\theta,X) &=& N_{\rm max}\times\exp((X_{\rm max}-X)/\lambda)\\
\nonumber
&\times&\left(\frac{X-X_0}{X_{\rm max}-X_0}\right)^{(X_{\rm max}-X_0)/\lambda}\times\frac{X_{\rm max}-X}{\lambda(X-X_0)}\\ \nonumber
&\times&F(E,E_\mu,\theta,X)\times\frac{1}{fE_\mu\cos\theta X}\times\left(1-\frac{AE_\mu}{E_0}\right)^{\alpha_2}. \nonumber
\end{eqnarray}
The first two lines on the right side of \refeq{eq:formula} are the derivative of the 
G-H formula. Because the application here is to hadronic cascades, the values of the parameters for number of particles at shower maximum ($N_{max}$), depth of shower maximum ($X_{max}$), starting depth ($X_0$) and interaction scale ($\lambda$) are numerically quite different from those of the original G-H formula for air showers dominated by the electromagnetic cascade. In the last line, the first two factors give
the probability of decay of a meson of energy $E$
to a muon of energy $E_\mu$ relative 
to the total rate of decay and re-interaction.  The last factor is the threshold factor
as in the Elbert formula.  In the denominator of the decay factor,
$E_\mu$ is replaced by $fE_\mu$ where
the factor $f$ represents the ratio between the minimum muon energy and the mean energy of muons above this threshold. Its behaviour is inferred from simulations (described below) and can be seen in \reffig{fig:factor}. There is an increase from 1 at threshold up to a value $f \approx 2.45$ at $E_0/A \gtrsim 10^{2.72} E_\mu$, after which it remains constant. The $\sec\theta$ factor reflects the
increase of the muon flux with zenith angle, and the $1/X$ dependence reflects the
importance of high altitude (low density) for decay. 

We consider two channels for muon production, $\pi^\pm\rightarrow \mu + \nu_\mu$ 
(100\% branching ratio) and decay of charged kaons (63.5\%).  The decay fraction for
charged pions with interaction length $\lambda_\pi$ is of the form 
$$\frac{1/d_\pi}{1/d_\pi + 1/\lambda_\pi},$$ where 
$$\frac{1}{d_\pi}=\frac{\epsilon_\pi}{E_\pi \cos\theta X}$$ and the pion critical
energy is given by
\begin{equation}
\epsilon_\pi=\frac{m_\pi c^2}{c\tau_\pi}\frac{RT}{Mg}
\approx \SI{115}{\GeV}\times\frac{T}{\SI{220}{\kelvin}},
\label{eq:critical_energy}
\end{equation}
where $m_\pi$ and $\tau_\pi$ are the mass and lifetime of the pion, $g$ is the gravitational constant, $R$ is the molar gas constant, $M = \SI{0.028964}{\kg \per \mol}$ for dry air and $T$ is the temperature.
The critical energy for charged kaons is larger by a factor of $7.45$ corresponding
to its larger mass and shorter decay length.
In pion decay, the muon carries an average energy of $E_\mu=r_\pi\times E_\pi$,
with $r_\pi\approx 0.79$.
The corresponding factor for decay of charged kaons has $E_\mu=r_K\times E_K$ with
$r_K\approx 0.52$.  In \refeq{eq:formula} the common factor
$1/fE_\mu\cos\theta X$ is factored out so that
\begin{equation}
F(E,E_\mu,\theta,X) = f_\pi\frac{r_\pi\epsilon_\pi\lambda_\pi}{1+\frac{r_\pi\epsilon_\pi\lambda_\pi}{fE_\mu\cos\theta X}}
+f_K\frac{r_K\epsilon_K\lambda_K}{1+\frac{r_K\epsilon_K\lambda_K}{fE_\mu\cos\theta X}}.
\label{eq:Fdecay}
\end{equation} 
In this equation, $f_\pi=0.92$ and $f_K=0.08$ are the relative fractions of momentum carried by charged pions and charged kaons after accounting for the branching ratio $0.635$ for kaon decay to muons.  The numerical values are based on Fig.~5.2 of \refref{Gaisser:2016uoy} where the momentum fraction carried by charged pions in p-air interactions is $Z_{N,\pi}(\gamma=1) = 0.29$, and the fraction carried by charged kaons is $0.040$.  So $f_\pi=0.29/(0.29+0.635\times0.04)=0.92$, and $f_K=1-f_\pi = 0.08$.

\begin{figure}
\includegraphics[width=0.6\textwidth]{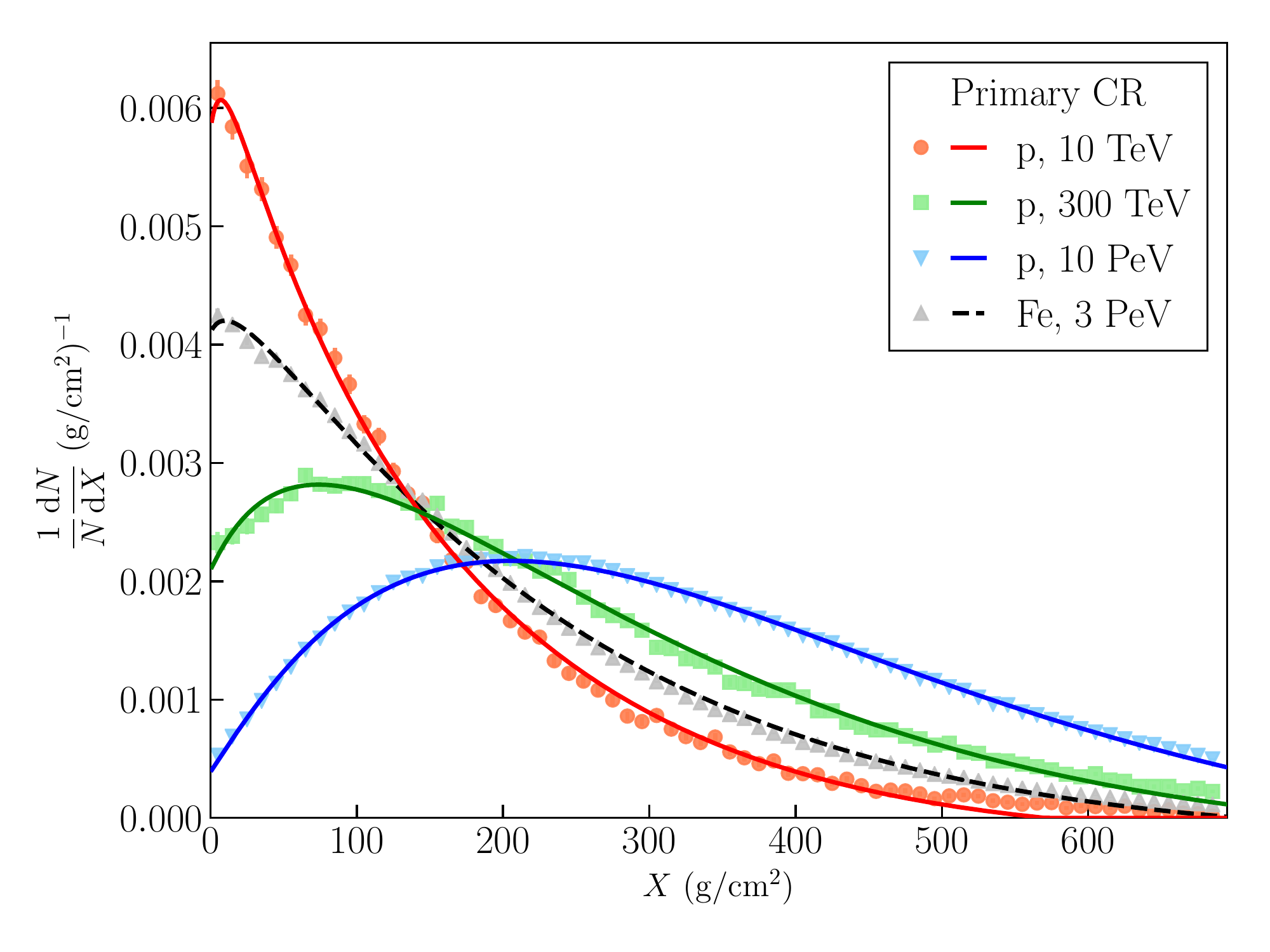}
\centering
\caption{Differential muon production spectrum normalized to total $N_\mu$ for  
$E_\mu\ge \SI{300}{\GeV}$ for different vertical primary cosmic rays. The markers are values obtained from CORSIKA simulations, the solid lines are the result of fitting \refeq{eq:formula} to these values.}
\label{fig:profiles}
\end{figure}

\begin{figure}
\centering
\includegraphics[width=.7\textwidth]{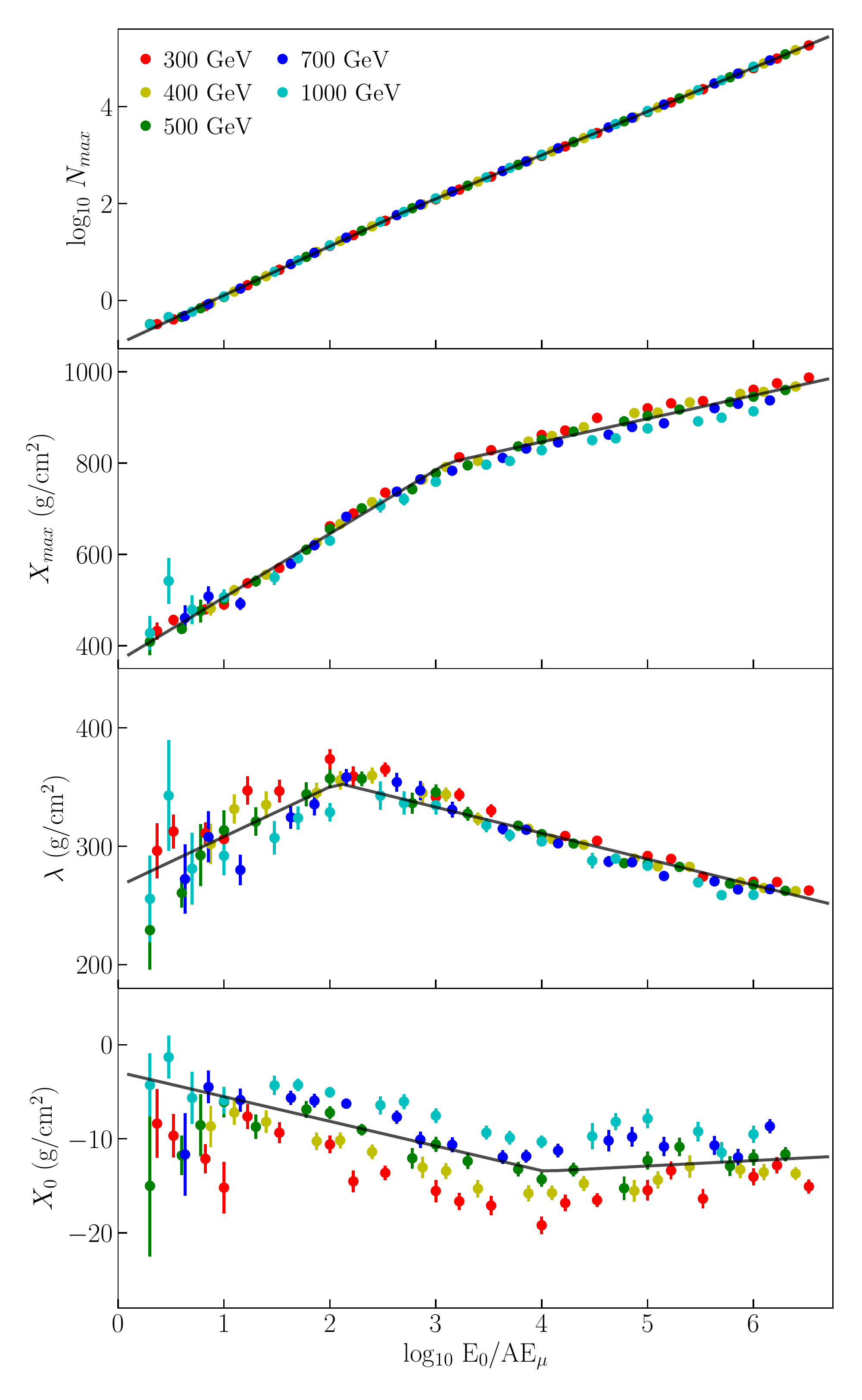}
\caption{Optimal values of the four parameters in fits of the production spectrum \refeq{eq:formula} to simulations of vertical proton showers with different threshold muon energies over a large range of primary energies. The black line shows the approximate description given by \refeq{eq:params} and \reftab{tab:params}.}
\label{fig:functions}
\end{figure}

The muon production spectrum of \refeq{eq:formula} is fitted to simulations produced with
CORSIKA~\cite{Heck:1998vt} v7.7100 using Sibyll2.3c~\cite{Engel:2019dsg} as the high-energy interaction model and UrQMD~\cite{Bass:1998ca, Bleicher:1999xi} for interactions below \SI{80}{\GeV}, relevant for \refsec{sec:rates}. An atmospheric profile corresponding to the average April South Pole atmosphere between 2007 and 2011~\cite{DeRidder:2019} was used. The fit of the formula to a muon production profile obtained from simulations has four free parameters, $N_{\rm max}$, $\lambda$, $X_{\rm max}$ and $X_0$, in the derivative of the G-H function. 
\reffig{fig:profiles} shows examples of fitted longitudinal profiles of muon production
over a range of primary energies for \SI{300}{\GeV}, normalized to the
total number of muons produced for each primary energy and mass. 
Because the simulation includes production of muons from all channels, the contribution 
of kaons is included implicitly.
The resulting optimal values of $N_{\rm max}$, $\lambda$, $X_{\rm max}$ and $X_0$ from repeating this procedure for a large number of muon and primary cosmic ray energies can be seen in \reffig{fig:functions}, where it becomes clear that they depend (in leading order) on $E_0/AE_\mu$. This behaviour is fit with the following functions,
\begin{eqnarray}
\label{eq:params}
N_{\rm max}&=& c_i\times A \times \left(\frac{E_0}{AE_\mu}\right)^{p_i}\\ \nonumber
X_{\rm max}, \lambda, X_0&=&a_i + b_i\times \log_{10}\left(\frac{E_0}{AE_\mu}\right),\\ \nonumber
\end{eqnarray} 
where $c_i$, $p_i$, $a_i$, and $b_i$ are defined for each function separately and have two regimes with a break
at $R_b = \frac{E_0}{AE_\mu} = 10^{q}$ and parameters $(a_i,b_i)$ with $i=1$
below the break and $i=2$ above.  All parameters are listed in \reftab{tab:params}
for results based on simulations of vertical showers with $E_\mu > 300, 400, 500, 700, \SI{1000}{\GeV}$.  As noted in \refref{Song:2004pk}, the parameter $X_0$
is often negative in fitting individual showers and is to be considered simply as a
parameter rather than the starting point of the cascade. For depths $X > X_{max}$, \refeq{eq:formula} gives negative values, in which case we set it to zero.

\begin{table}[htb]
\begin{center}
\caption{Parameter values for \refeq{eq:params} for $\SI{300}{\GeV} \lesssim E_\mu \lesssim \SI{1}{\tera \eV}$.}
\begin{tabular}{lr|r|r|r}
\hline \hline
&$i$& $c_i$ & $p_i$ & $q$\\
\hline \hline
$N_{\rm max}$ &1&0.124&1.012&2.677 \\
&2&0.244&0.902& \\
\hline \hline
& $i$ & $a_i$ (g/cm$^2$)& $b_i$ (g/cm$^2$)& $q$ \\
\hline \hline
$X_{\rm max}$ &1& 366.2 & 139.5 & 3.117 \\
&2& 642.2 & 51.0 & \\
\hline
$\lambda$ &1& 266.0 & 42.1 & 2.074 \\
&2& 398.8 & -21.9 & \\
\hline
$X_0$ & 1& -2.9 & -2.6 & 4.025 \\
&2& -15.8 & 0.6 & \\
\hline
$f$ & 1 & 1 & 0.53 & 2.72 \\
& 2 & 2.45 & - \\
\hline \hline
\refeq{eq:ElbertFormula}: && $K = 12.4$ & $\alpha_1 = 0.787$ & $\alpha_2 = 5.99$ \\
\hline\hline 
\end{tabular}
\label{tab:params}
\end{center}
\end{table}


\begin{figure}
\includegraphics[width=0.6\textwidth]{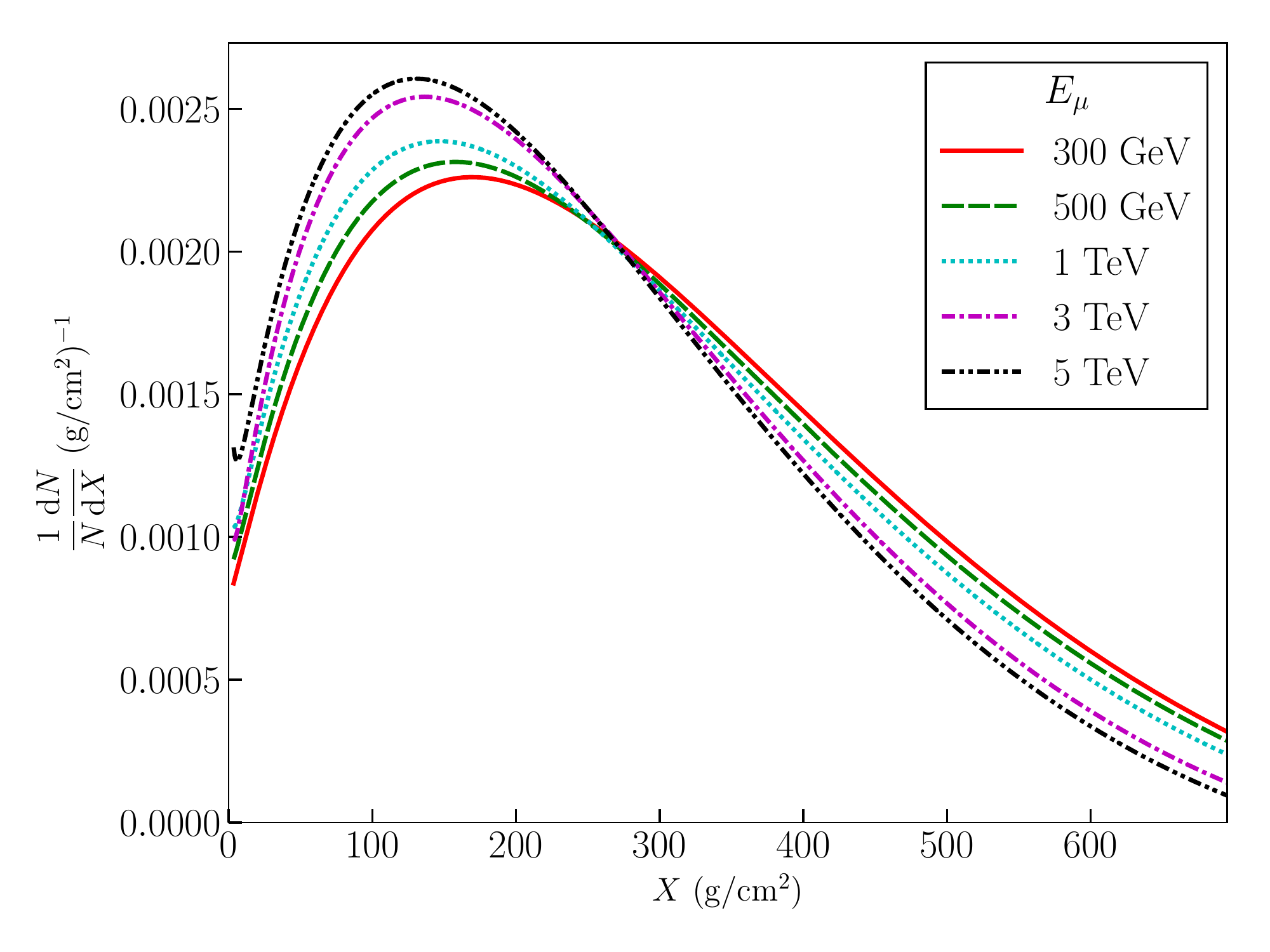}
\centering
\caption{Normalized muon production profiles fit to vertical proton simulations with primary energies chosen so that $E_0/E_\mu = 10^4$ for each threshold muon energy. There is a clear remaining $E_\mu$ dependence in the shapes, with production for higher energy muons peaking earlier in the atmosphere.}
\label{fig:scalingviolation}
\end{figure}

\begin{figure}
\includegraphics[width=0.5\textwidth]{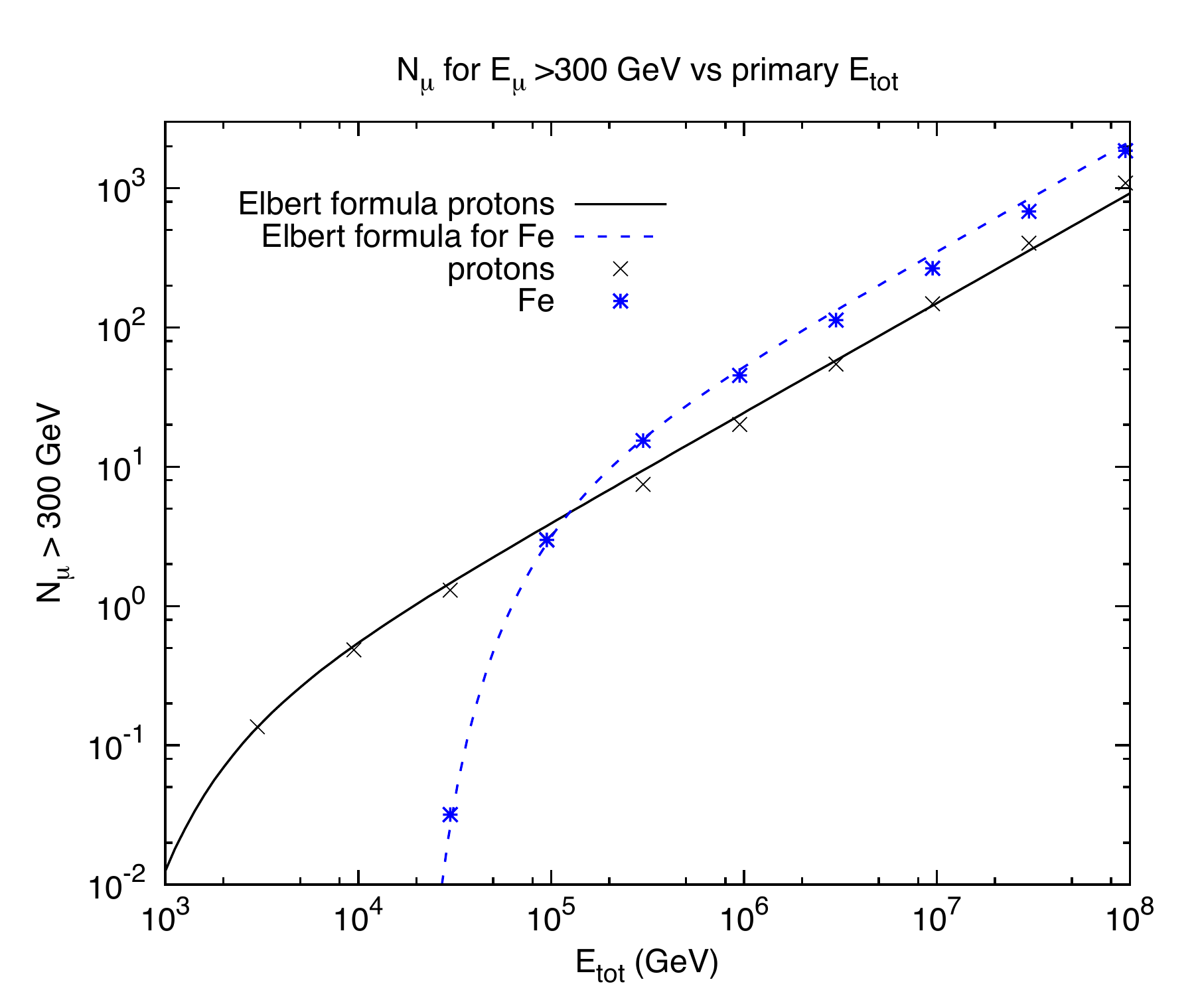}\includegraphics[width=0.5\textwidth]{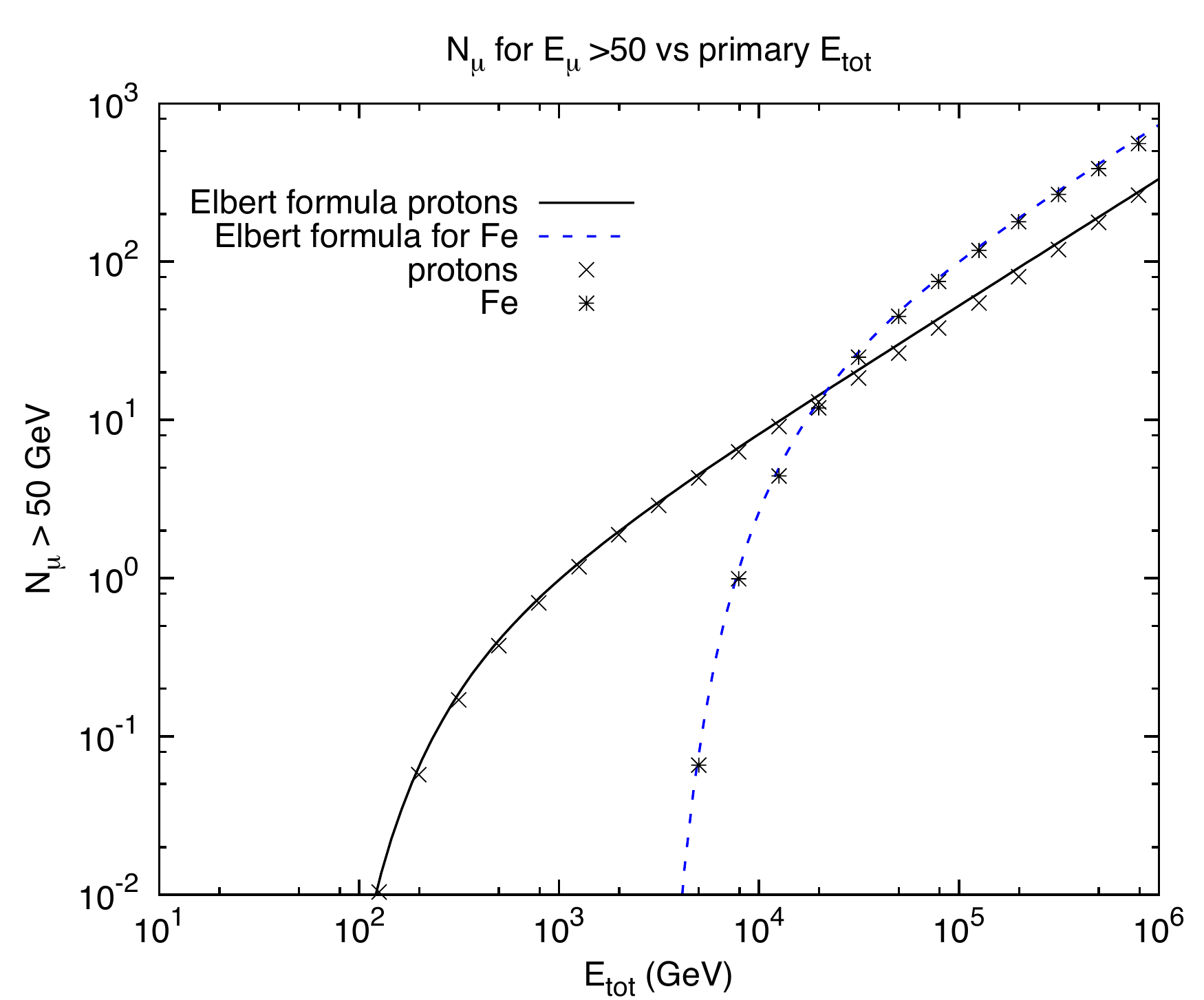}
\caption{$N_\mu > E_{\mu,{\rm min}}$ for vertical protons (solid) and Fe (dashed) from \refeq{eq:ElbertFormula} 
compared to values from the integration of \refeq{eq:formula} for two energy regimes corresponding to Tables~\ref{tab:params} and \ref{tab:params_ND}.}
\label{fig:Nmu_vs_Ea}
\end{figure}

As can already be seen in \reffig{fig:functions}, the muon production profiles do not scale perfectly with the ratio of primary energy per nucleon and minimum muon energy, but show a remaining dependency on the latter. This is further illustrated in \reffig{fig:scalingviolation}, where it becomes clear in the normalized production profiles that the muon production peaks higher in the atmosphere for higher $E_\mu$. For this reason, it is best to optimize the parameterization using simulations for a specific energy range for the application at hand. In \reftab{tab:params_ND}, we give a set of values fitted to simulations with a muon threshold of \SI{50}{\GeV}, which is used for calculations for NOvA Near Detector~\cite{Acero:2019lmp} in \refsec{sec:rates}. For this case, also the behaviour of the $f$ ratio is different and it now has two regimes with a clear energy dependence instead of becoming constant at high $E_0/AE_\mu$.
In \reffig{fig:Nmu_vs_Ea}, the muon multiplicities integrated over depth are 
compared with the standard Elbert formula for both energy regions.

A Python implementation of the parameterizations derived in this work can be found on GitHub\footnote{\url{https://github.com/verpoest/muon-profile-parameterization}}.

\section{Surface-underground coincident measurements}
\label{sec:surface-underground}

Before applying the parameterization in the context of underground muon rates, it is instructive to look at the case where the primary cosmic-ray energy is fixed, as can be achieved with coincident measurements of air showers with a surface array and an underground muon detector. Examples of experiments with this capacity are EAS-TOP and MACRO at Gran Sasso~\cite{Aglietta:2004ws}, the Baksan Underground Laboratory~\cite{Bakatanov:1999xp}, SPASE-AMANDA~\cite{Ahrens:2004dg,Ahrens:2004nn} and the IceTop and IceCube detectors at the South Pole~\cite{IceCube:2019hmk}. We will use primary and muon energies relevant for the latter in our calculations.

As with the Elbert formula \refeq{eq:ElbertFormula}, it is possible to estimate the multiplicity of high-energy muons produced in a shower given a certain primary energy, mass and muon energy threshold by integrating over the production profile given by \refeq{eq:formula} and the parameterization of e.g. \reftab{tab:params}. More information about the muon bundle can however be extracted from the profiles, as the depth dependence of the muon production combined with a transverse momentum distribution allows one to estimate its lateral size. Furthermore, the dependence on the atmospheric temperature through the critical energy \refeq{eq:critical_energy} for pions and kaons, enables the calculation of how both the muon bundle multiplicity and size vary throughout the year.

\begin{figure}
     \centering
     \begin{subfigure}[b]{0.45\textwidth}
         \centering
         \includegraphics[width=\textwidth]{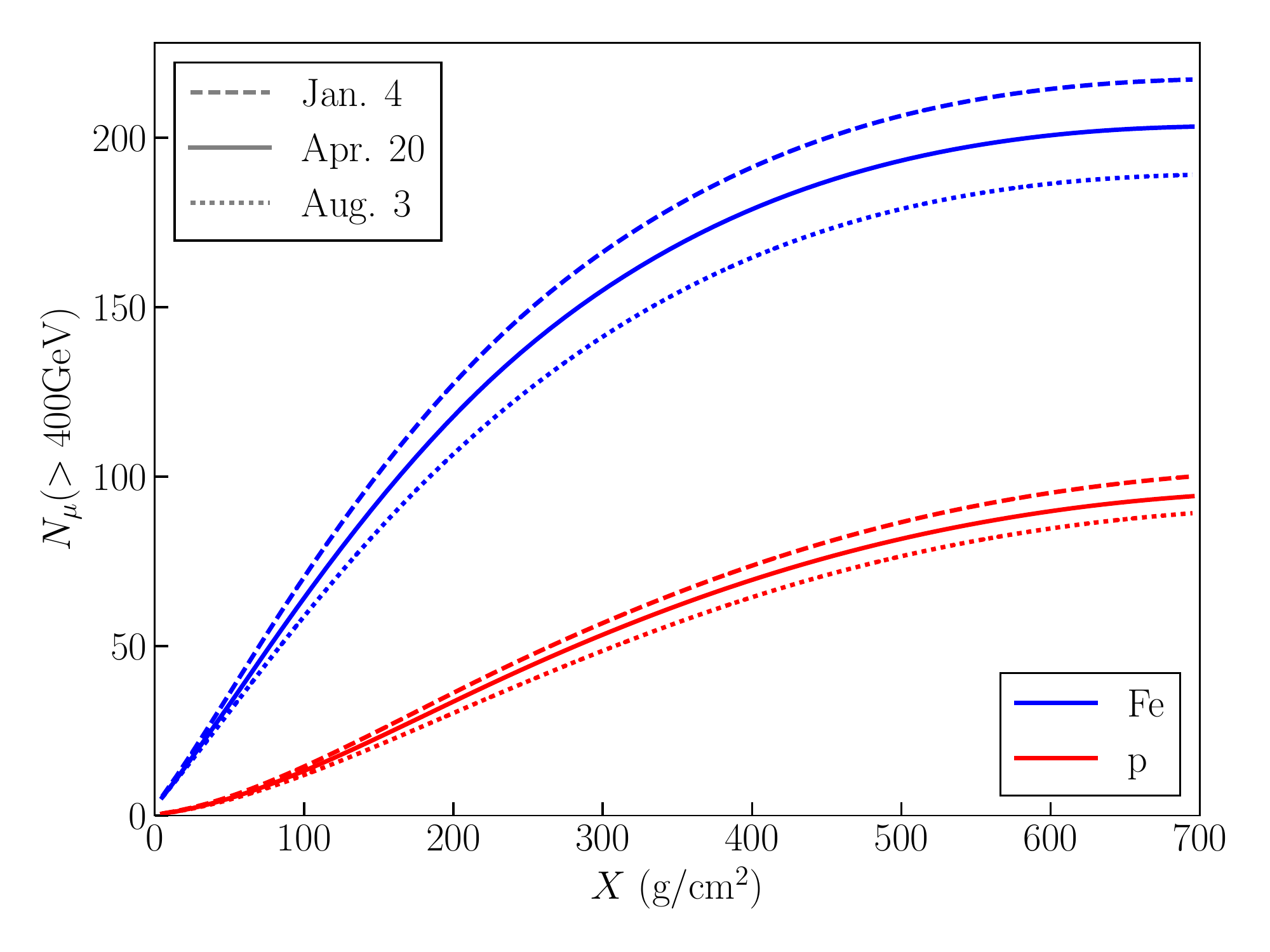}
         \caption{}
         \label{fig:multiplicity_var_a}
     \end{subfigure}
     \hfill
     \begin{subfigure}[b]{0.45\textwidth}
         \centering
         \includegraphics[width=\textwidth]{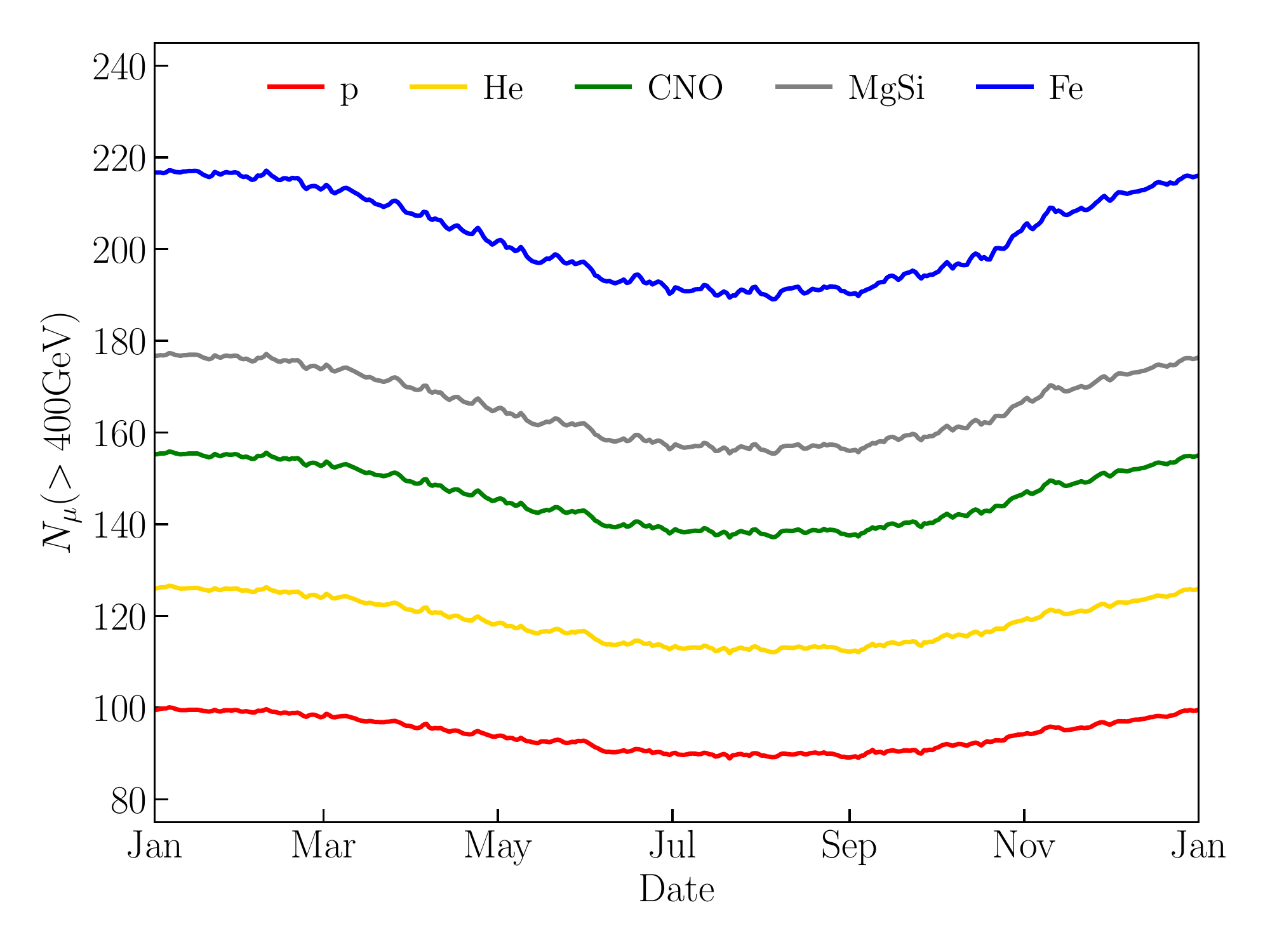}
         \caption{}
         \label{fig:multiplicity_var_b}
     \end{subfigure}
        \caption{Effect of atmospheric variations on high-energy muon production in air showers. (a) Integral muon number throughout the atmosphere at the South Pole for vertical \SI{10}{\peta \eV} proton and iron primaries for three different days of 2017, corresponding to days with approximately minimum, maximum and mean muon multiplicities. (b) Variation of the expected muon multiplicity for five different mass groups throughout the year.}
        \label{fig:multiplicity_var}
\end{figure}

We estimate the seasonal variations of the multiplicity as relevant for IceCube, which measures muons with energies above approximately \SI{400}{\GeV} from primary cosmic rays in the energy range \SI{1}{\peta \eV} to \SI{1}{\exa \eV}. Atmospheric data is obtained from the AIRS satellite~\cite{NASA:2018abc}, which provides the temperature at different atmospheric pressure levels unevenly spaced between \SI{1}{\hecto \Pa} and \SI{700}{\hecto \Pa}, between which we interpolate. Using the relation between vertical depth and pressure from the atmospheric overburden, the temperature as function of depth is obtained, which is used to calculate the muon production profiles. The resulting integral profiles can be seen for vertical showers of \SI{10}{\peta \eV} primaries in \reffig{fig:multiplicity_var_a} for three different days representing roughly the minimal, maximal and mean expected muon number in a year. During the austral summer, the atmosphere is warmer and less dense, causing more mesons to decay to high-energy muons instead of undergoing further interaction, and vice versa for the winter. This can clearly be seen in \reffig{fig:multiplicity_var_b}, where the variation of the expected muon number throughout the year is shown for 5 different cosmic-ray mass groups. The maximal relative variations are of the order of 6\% around the mean, slightly increasing with primary mass, and slowly decreasing with primary energy. This may be a relevant effect to take into account when deriving primary cosmic-ray composition from this observable by comparing experimental data to simulation. Variations of this kind have already been observed and were corrected for in \refref{DeRidder:2013way, IceCube:2019hmk}.

\begin{figure}
     \centering
     \begin{subfigure}[b]{0.45\textwidth}
         \centering
         \includegraphics[width=\textwidth]{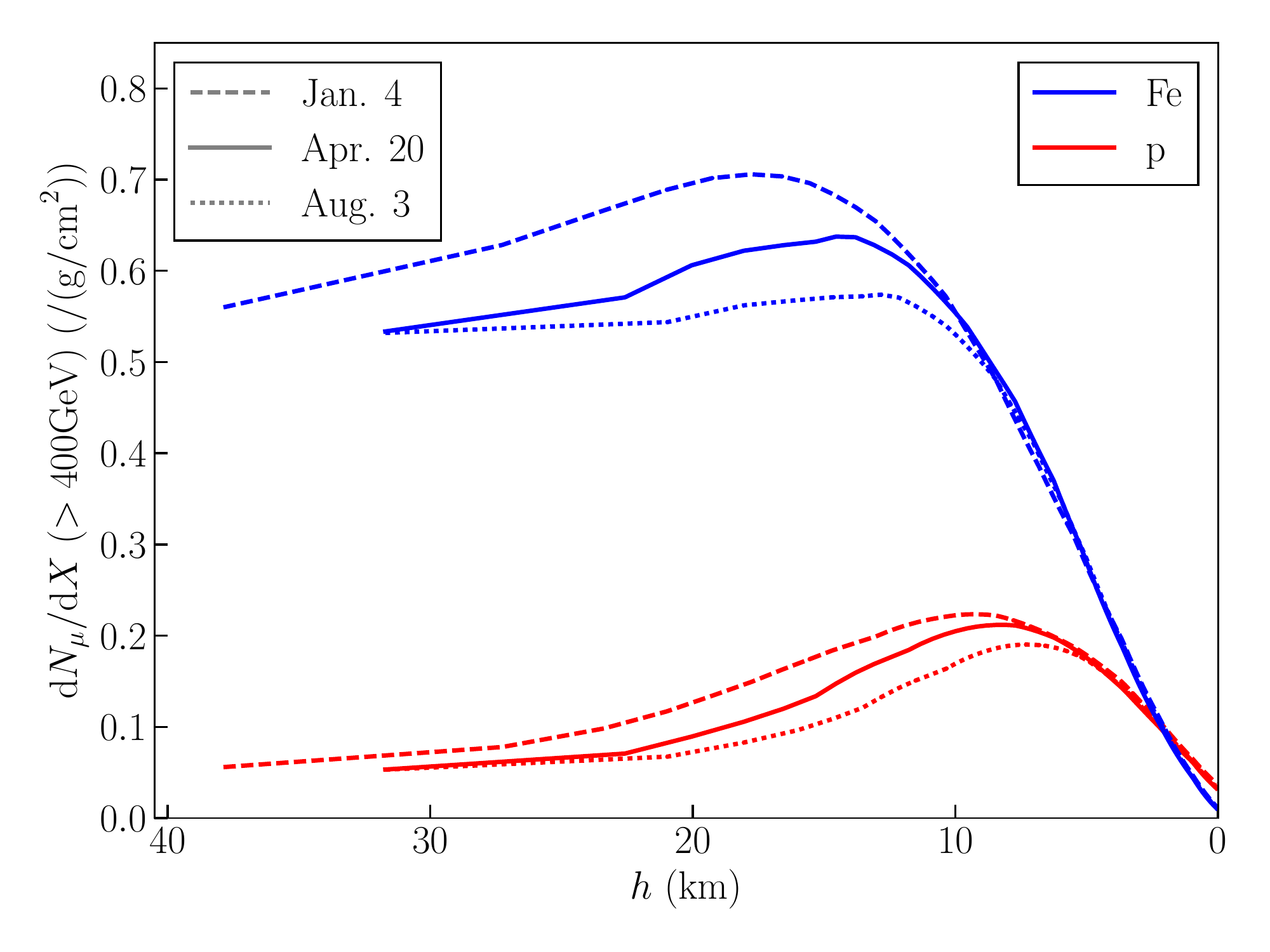}
         \caption{}
         \label{fig:size_var_a}
     \end{subfigure}
     \hfill
     \begin{subfigure}[b]{0.45\textwidth}
         \centering
         \includegraphics[width=\textwidth]{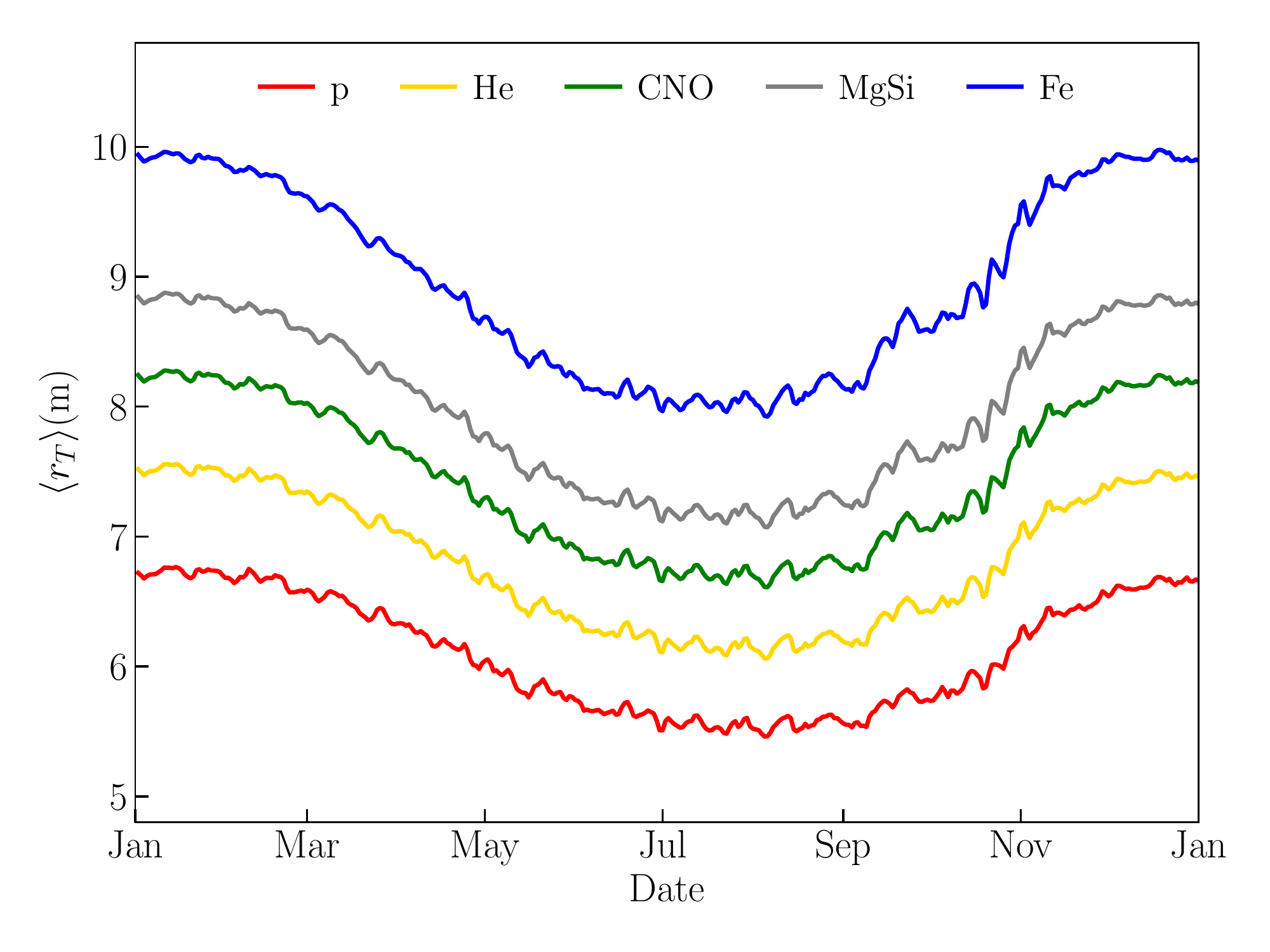}
         \caption{}
         \label{fig:size_var_b}
     \end{subfigure}
        \caption{Effect of atmospheric variations on high-energy muon bundle size.  (a) Muon production as function of height above the surface for $\SI{10}{\peta \eV}$ proton and iron primaries at the South Pole on three different days of 2017, corresponding to days with approximately minimum, maximum and mean muon multiplicities. (b) Variation of the mean bundle size for five different mass groups throughout the year.}
        \label{fig:size_var}
\end{figure}

To estimate the transverse size of the muon bundle, the muon production as function of altitude rather than atmospheric depth is important, and needs to be combined with a transverse momentum distribution for the muons. The transverse distance from the shower axis of a muon with energy $E_\mu$ produced at an altitude $h$ is given by 
\begin{equation}
    r_T = \frac{p_T}{E_\mu} \times \frac{h}{\cos \theta},
    \label{eq:transverse_dist}
\end{equation}
where $\theta$ is the zenith angle of the primary and $p_T$ is the transverse momentum of the muon. The altitude corresponding to the vertical depth $X_v$ where the temperature is measured, is calculated using the ideal gas law, which gives

\begin{equation}
    h(X_v) = \frac{RT}{Mg} \ln \frac{X_0}{X_v},
    \label{eq:altitude}
\end{equation}
with $X_0$ the vertical depth at $h=0$. A transverse momentum distribution $\mathrm{d}\sigma/\mathrm{d}p_T^2 = f(p_T)$ can be assumed for the muons, relative to the direction of the incident primary particle, with the normalized distribution given by
\begin{equation}
    f(p_T) = \frac{4p_T}{\langle p_T \rangle^2} e^{-2p_T/\langle p_T \rangle},
    \label{eq:p_T}
\end{equation}
where $\langle p_T \rangle \approx \SI{350}{\mega \eV}$~\cite{Alper:1974me,Alper:1974xp}.
We perform a simple estimate of the expected bundle size using this mean value of the transverse momentum for the surface above IceCube, located at an elevation of \SI{2835}{\m} with an atmospheric depth of about \SI{700}{\gram \per \cm \squared}, which is used as the zero point for our altitude calculations. \reffig{fig:size_var_a} shows the muon production differential in depth as function of altitude for proton and iron primaries and at three dates again representing roughly an average and two extremal days. It is clear that muon production happens higher up in the atmosphere for heavy primaries. For a specific primary mass, muons are also produced higher in the atmosphere in summer compared to winter, because of its thermal expansion. Calculating the transverse distance for a muon with an average $p_T$ produced at a depth $X$ using \refeq{eq:transverse_dist}, and taking the weighted average by multiplying with the production profile and integrating over depth, gives an estimate of the expected bundle radius for muons with energy $> E_\mu$, in this case \SI{400}{\GeV}. Results of this calculation for various primary masses over a full year are shown in \reffig{fig:size_var_b}. The summer maximum caused by the change in production altitude can clearly be observed. The seasonal variations in the size are of the order of 10\%. One can also calculate the size for muons with the average muon energy in the shower instead of the minimum energy by multiplying $E_\mu$ in \refeq{eq:transverse_dist} by the $f$ ratio defined in the previous section. 

A full calculation of the muon bundle properties in the detector would need to include the propagation to the detector, where multiple Coulomb scattering~\cite{Lipari:1991ut} in the overburden would further separate the muons, as well as charge separation of the muons caused by the geomagnetic field~\cite{Abreu:2011ki}, but is beyond the scope of this example.

The simple calculation using the parameterization predicts a muon bundle size that is about 40\% larger for iron primaries than for protons. The amount of muons expected for iron primaries is, on the other hand, roughly a factor of 2 higher than for protons. A similar conclusion was reached as a result of simulations in \refref{Forti:1990st}.

\section{Rates of muons in underground detectors}\label{sec:rates}
The rate of muons of energy $>E_\mu$ from a direction corresponding to zenith angle $\theta$ 
in a detector with area $A_{\mathrm{eff}}$ is given by
\begin{equation}
I_\mu(>E_{\mu,\mathrm{min}},\theta) = A_{\mathrm{eff}}(E_{\mu,\mathrm{min}},\theta)\int{\rm d}X\int_{E_{\mu,\mathrm{min}},\theta}{\rm d}E_\mu\, P(E_\mu,\theta,X),
\label{eq:rtheta}
\end{equation}
where now
\begin{equation}
P(E_\mu,\theta,X)=\int_{E_{\mu,\mathrm{min}},\theta}\frac{{\rm d}N}{{\rm d}X}(>E_{\mu,\mathrm{min}},E_0,A=1,\theta,X)\,\phi(E_0){\rm d}E_0
\label{eq:PofEmu}
\end{equation}
 is the production spectrum of muons differential in slant depth $X$
 from \refeq{eq:formula} folded with the primary spectrum.  We use the superposition approximation with $\phi(E_0)$ the spectrum of nucleons per GeV/nucleon summed over all nuclei
 using the H3A model~\cite{Gaisser:2011cc} to estimate rates of events.

In this section, we consider three compact underground detectors, the MINOS Far Detector (FD) at Soudan~\cite{Adamson:2009zf}, the MINOS Near Detector (ND) at Fermilab~\cite{Adamson:2014xga} and the NOvA ND~\cite{Acero:2019lmp}, also at Fermilab.  For these detectors, 
all muons with energy greater than the minimum to reach the detector can be counted and the same energy threshold applies to all muons in an event from a given direction.  
The MINOS FD is at a depth corresponding to a minimum muon energy of \SI{730}{\GeV}, which increases with zenith according to the energy-loss  for each slant depth.  In contrast, the shallow MINOS ND and the NOvA ND have thresholds around \SI{50}{\GeV}.  In this case we account for muon decay and energy loss in the atmosphere, which reduces the rate by a few per cent.  In both cases the rates are calculated and summed over 8~bins of zenith angles with threshold muon energies given in \reftab{tab:Emin}.
\begin{table}[htb]
   \caption{Minimum muon energies (GeV) for 8 bins of $\cos\theta$}
    \centering
    \begin{tabular}{r|c|c|c|c|c|c|c|c} \hline
      $\cos\theta$&0.95&0.85&0.75&0.65&0.55&0.45&0.35&0.25  \\ \hline
         MINOS FD&  730&850&1030&1320&1800&2730&5000&14000\\ \hline
         NOvA ND&50&56&64&74&89&111&147&217 \\ \hline
    \end{tabular}
     \label{tab:Emin}
\end{table}

 The temperature dependence of the muon production spectrum is entirely contained
in the critical energies given in \refeq{eq:critical_energy}.
The positive correlation of the overall rate with temperature reflects
the higher probability of decay of the parent mesons to muons compared to re-interaction when the density is lower.
Event rates are generally higher in summer and lower in winter.
The correlation coefficient is larger for the
deeper MINOS FD~\cite{Adamson:2009zf}  than for the shallow detectors.  This is because, at the lower energy
threshold, pions still have a high probability of decaying, so the
correlation of the charged pion decay channel is only weakly coupled
to temperature.  For quantitative estimates of correlation of rates with effective temperature in
this section we use the temperature profiles from the AIRS satellite data~\cite{NASA:2018abc} 
at the locations of Soudan and Fermilab to calculate $T_{\rm eff}$.
The effective temperature is evaluated as
\begin{equation}
T_{\mathrm{eff}}(\theta)=\frac{\int{\rm d}X\,P(E_\mu,\theta,X)\,T(X)}{\int{\rm d}X\,P(E_\mu,\theta,X)},
    \label{eq:Teff}
\end{equation}
with $P(E_\mu,\theta,X)$ defined in \refeq{eq:PofEmu}.~\footnote{A different expression that uses the derivative with respect to temperature of the muon production spectrum~\cite{Grashorn:2009ey} is used by MINOS and other underground detectors.  We note the relatively small difference of using that definition in the analysis of the MINOS FD below.}
In Eq.~\ref{eq:Teff} $A_{\rm eff}(E_{\mu,{\rm min}})$ factors out of the integrals and cancels in the ratio.

Both MINOS detectors~\cite{Adamson:2015qua} and the NOvA ND~\cite{Acero:2019lmp} report anti-correlation with
 $T_{\rm eff}$ in the rates of events with two or more muons.  This is in contrast with
the seasonal variation of the total rate, which peaks in the summer when temperatures are higher.
Our goal here is to use the parameterization to estimate the extent to which the anti-correlation
for multiple muons can be accounted for by the effect of larger muon bundle radius in summer when
muon production occurs higher in the atmosphere (see \reffig{fig:size_var_a}).

The mean perpendicular distance from the shower axis of a muon produced at altitude $h$
with zenith angle $\theta$ is given by \refeq{eq:transverse_dist}.  To calculate the distance
from the shower axis at the level of the detector, $r_T(X)$ is weighted with the
production spectrum for each angular bin with slant depth $X$ related to altitude
 by \refeq{eq:altitude}.
%
\begin{equation}
\langle r_T(\theta)\rangle = \frac{\int {\rm d}E_{\mu}\,\int dX\,r_T(X)\,P_{\mu}(E_{\mu}, \theta, X)}{\int {\rm d}E_{\mu}\,\int dX\,P_{\mu}(E_{\mu}, \theta,X)}.
\label{eq:reff}
\end{equation}
Finally, $\langle r_T(\theta)\rangle$ is integrated over zenith angle.

There are two other processes that affect the separation of muons from the shower axis and from each other: multiple scattering in the overburden and separation of muons by bending in the geomagnetic field before they reach the surface.  Only the latter is important for the shallow detectors.  We first analyse the altitude effect and then comment on multiple scattering and magnetic deviations.

\subsection{The MINOS Far Detector at Soudan}\label{subsec:TeV}
The MINOS FD is at a depth of 2100 m.w.e., which corresponds to $E_{\mu,{\rm min}}\approx \SI{730}{\GeV}$.
In this case, the parameters of \reftab{tab:params} are appropriate.   
Integrating $r_T$ over production depth (altitude) and zenith angle, 
we find that the characteristic distance is
comparable to the \SI{8}{\m} lateral dimension of the detector.  For the assumed transverse momentum
distribution of \refeq{eq:p_T},
\begin{equation} 
{\rm Fraction} = 1 - \left(1+\frac{2p_T}{\langle p_T\rangle}\right)\times e^{-2p_T/\langle p_T\rangle}
\label{eq:fraction}
\end{equation}
gives the fraction of muons with momentum less than $p_T$.
The convolution of this fraction\footnote{Replace $r_T(X)$ in \refeq{eq:reff} with Fraction(X).}
with the muon production spectrum as a function of
slant depth (altitude) 
is calculated for three values of $p_{T,i}$ at each depth $i$ corresponding to $r_{T} = 8$, $4.5$ and \SI{0.6}{\m} at the ground.
\refeq{eq:p_T} is used to find $p_{T,i}$ at each depth $i$ corresponding to the three values of separation
that characterize multiple muon events in the MINOS FD.  In this way the partial rates for 
regions A ($0.6<r<\SI{4.5}{\m}$), B ($4.5<r<\SI{8}{\m}$) and C ($r>\SI{8}{\m}$) are estimated.  The idea for
this simple interpretation is that in an event with two or more muons, the closest muon to the shower
axis is a proxy for the center of the distribution, and the overall lateral distribution then
provides an estimate of the distance to other muons.
Approximating the MINOS FD as a cylinder of \SI{8}{\m} diameter and length \SI{31}{\m}, the total rate is estimated
as $\approx \SI{0.4}{\Hz}$.  The Elbert formula (\refeq{eq:ElbertFormula}) weighted by the primary
spectrum and assuming a Poisson multiplicity distribution is used to estimate the fraction of
events with two or more muons at the depth of the detector as $0.072$.  With these
assumptions, the predicted rate of multiple events is  \SI{0.028}{\Hz}, a factor of two larger
than observed.  This discrepancy is in part due to the fact that the 
shape of the detector is not accounted for in calculating the partial rates. 
For example, containment for events with two muons separated by $<\SI{8}{\m}$ would need 
to account for their orientation in azimuth relative to the long axis of the detector.

\begin{figure}
\includegraphics[width=\textwidth]{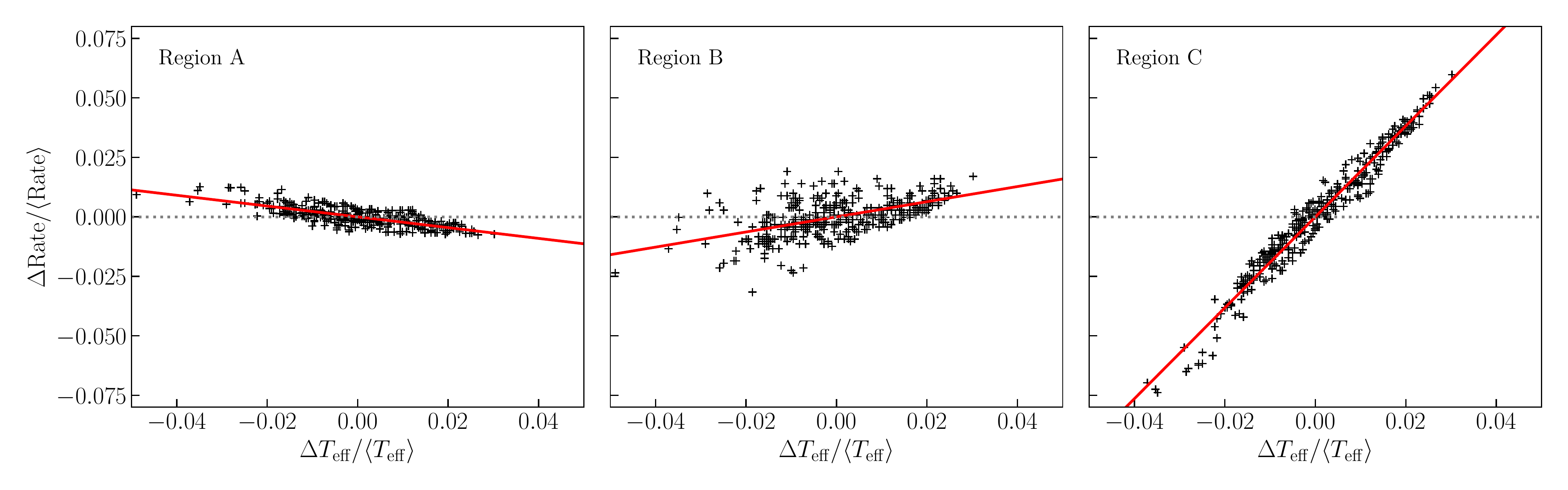}
\caption{Correlation of Regions A, B, and C in the MINOS Far Detector with $T_{\rm eff}$.  Partial correlation coefficients from the altitude effect only are
-0.23, 0.32 and 1.91 respectively for A, B, and C.}
\label{fig:ABCsf1}
\end{figure}

Our result is shown in \reffig{fig:ABCsf1} for calendar year 2009 by plots of the correlation between rates and effective temperature for the three regions.  The effect of multiple scattering is to add about $\approx \SI{2}{\m}$ to the separation of muons for deep detectors~\cite{Lipari:1991ut}.  The separation from propagation through the geomagnetic field~\cite{Abreu:2011ki} is $\approx \SI{1.5}{\m}$ for TeV muons.  To estimate the effects of multiple scattering and magnetic deflection, we therefore reduce the radii for the atmospheric effect from ($0.6, 4.5, 8$)~m to ($0.1, 2.5,6.)$~m.  Doing so increases the anti-correlation effect, changing the correlation coefficients for regions $(A, B, C) = (-0.23, 0.32, 1.91)$ to $(-0.46, 0.05, 1.54)$.  Using a definition of effective temperature similar to that of Ref.~\cite{Grashorn:2009ey} gives the values $(A,B,C) = (-0.51,0.09, 1.69)$ for the calculation with the reduced atmospheric effect.

The MINOS FD paper does not give explicit values for the correlation coefficients
for Regions A,B,C.  However, these can be inferred from Table~1 of \refref{Adamson:2015qua}, which gives the amplitudes and phases of
a sinusoidal fit to the rates in the three regions as well as corresponding
values for their earlier measurement of single muons in the MINOS~FD. The correlation coefficient for the single muons is 0.873~\cite{Adamson:2009zf} and the 
amplitude of the sinusoidal fit for single muons is 1.27\%.  The correlation
coefficients inferred in this way from the amplitudes and phases of regions A 
and C are respectively -0.69 and +1.38, with B intermediate.  The trend from negative correlation in region A to
strong positive correlation in Region~C is present in our calculation, and motivates further investigation accounting in detail for all three separation effects as well as detector geometry.

\begin{table}[htb]
\begin{center}
\caption{Parameter values for \refeq{eq:params} for $>\SI{50}{\GeV}$ muons.}
\begin{tabular}{lr|r|r|r}
\hline \hline
&$i$& $c_i$ & $p_i$ & q\\
\hline \hline
$N_{\rm max}$ &1&0.144&0.972&2.557 \\
&2&0.213&0.905& \\
\hline \hline
& $i$ & $a_i$ (g/cm$^2$)& $b_i$ (g/cm$^2$)& $q$ \\
\hline \hline
$X_{\rm max}$ &1& 260.9 & 176.4 & 3.476 \\
&2& 665.1 & 60.1 & \\
\hline
$\lambda$ &1& 289.4 & 95.0 & 1.526 \\
&2& 483.0 & -31.8 & \\
\hline
$X_0$ & 1& -28.7 & -2.3 & 2.778 \\
&2& -48.0 & 4.6 & \\
\hline
$f$ & 1 & 1 & 0.69 & 2.80 \\
& 2 & 3.06 & -0.05 \\
\hline \hline
\refeq{eq:ElbertFormula}: && $K = 6.034$ & $\alpha_1 = 0.80$ & $\alpha_2 = 5.99$ \\
\hline\hline
\end{tabular}
\label{tab:params_ND}
\end{center}
\end{table}

\subsection{Shallow detectors at Fermilab}\label{subsec:lowE}
For the MINOS and the NOvA ND,
both at $\approx 200$ m.w.e., where $E_{\mu,{\rm min}}\sim \SI{50}{\GeV}$, we use parameters of \reftab{tab:params_ND}, 
tuned for the lower energy region.  The main practical difference at the lower energy
is that the muon bundle size of $\approx \SI{60}{\m}$ is much larger than the scale of the detector.  In this case, the rate 
of multiple muon events is determined primarily by the small probability of a second muon to 
find the detector, as well as by the fraction of events with two or more muons at the depth.
Accordingly, we 
estimate the fraction of events with multiple muons by converting the transverse momentum variables in \refeq{eq:fraction} to lateral distances from the shower axis using the
weighted distribution of altitude/$\cos\theta$.  For a simple estimate we calculate the probability that a second muon lies within \SI{7}{\m} of the shower axis.  We find that 3.3\% of muons satisfy this condition.  Our calculated total rate in the NOvA ND with $E_{\mu,\mathrm{min}}=\SI{50}{\GeV}$ is \SI{50}{\Hz}.
Using the Elbert formula \refeq{eq:ElbertFormula} with the assumption of a Poisson distribution, we estimate that 16\% of events have two or more muons at the depth of NOvA. The product $\approx \SI{0.26}{\Hz}$ is somewhat greater than the observed rate of multiple muon events in Ref.~\cite{Acero:2019lmp}, which is \SI{0.15}{\Hz}.
 
\begin{figure}
\includegraphics[width=0.5\textwidth]{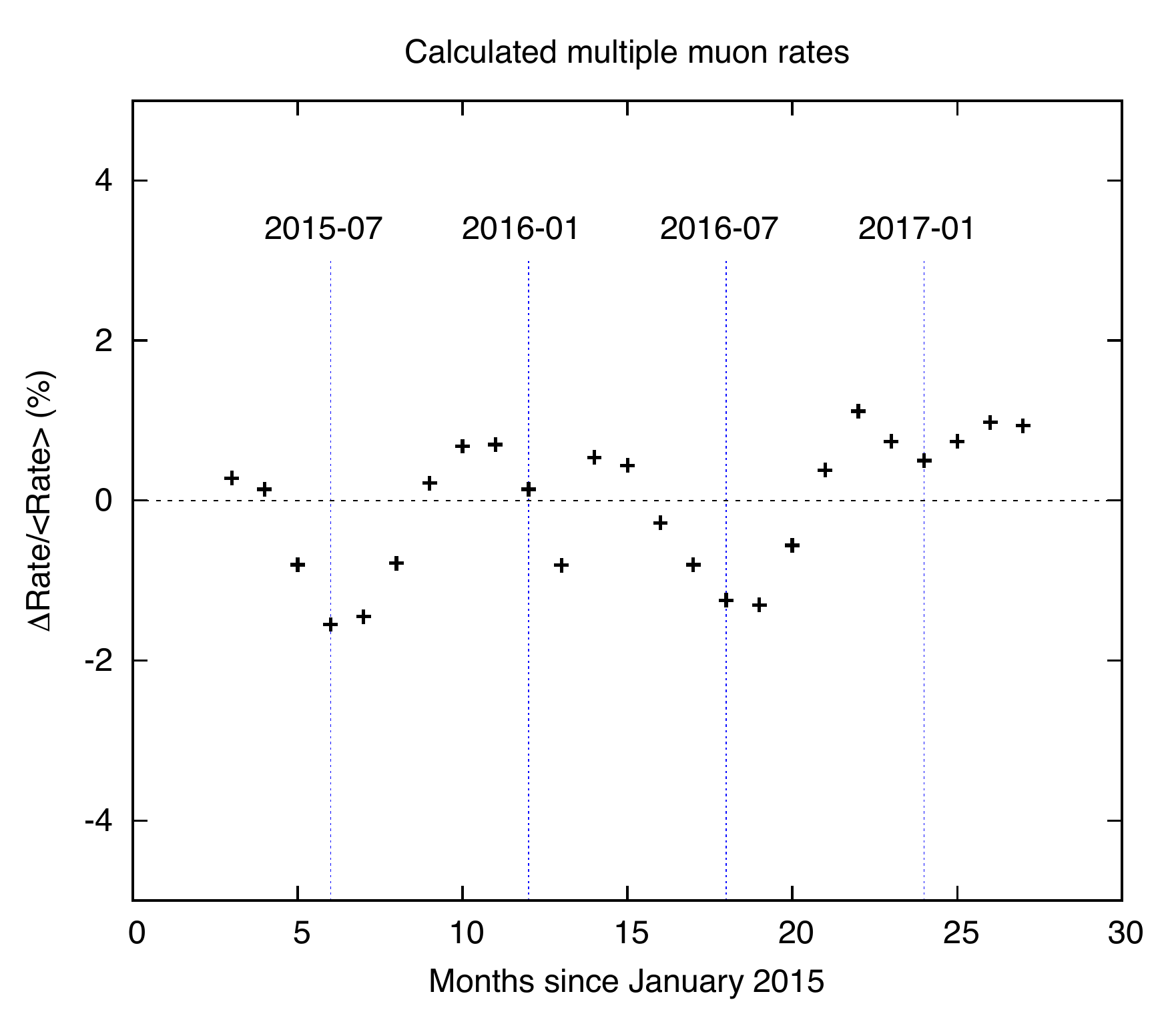}\includegraphics[width=0.5\textwidth]{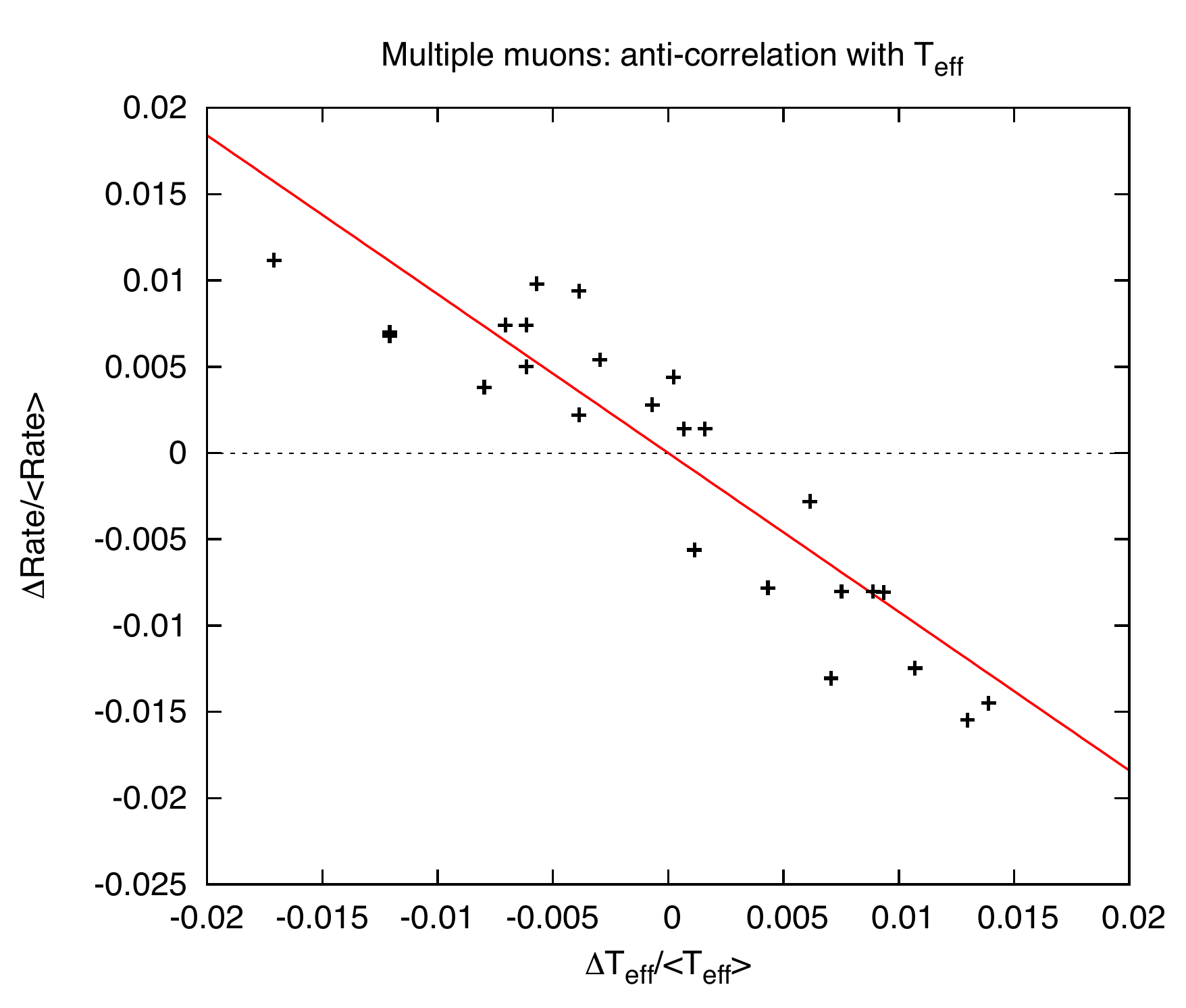}
\caption{Calculated rates (left) and correlation (right) for multiple muon events at NOvA.}
\label{fig:NOvA}
\end{figure}

 Results of our calculations for multiple muons at NOvA are shown in \reffig{fig:NOvA}.  The
 monthly averages show the anti-correlation with effective temperature as in Figs.~3 and 4 of
 \refref{Acero:2019lmp}, but the amplitude of the variation is smaller.  The right panel of \reffig{fig:NOvA}
 shows a correlation coefficient of $\sim -0.92$, as compared to $-4.14$ reported in \refref{Acero:2019lmp}.  The separation from propagation in the geomagnetic field for $\sim \SI{50}{\GeV}$ muons at Fermilab is only somewhat less than the \SI{60}{\m} scale of the altitude effect.  To estimate the geomagnetic effect, we reduce the parameter of the altitude effect from $7$ to \SI{4}{\m}.  This gives only a small increase in the anti-correlation effect, changing the correlation coefficient from $-0.92$ to $-1.00$.
In summary, the analysis described here does not fully explain the observed anti-correlation in NOvA, 
indicating that a full simulation is required.

\section{Summary}
The main point of this paper is to provide a parameterization of the muon production
profile as a function of atmospheric depth and zenith angle for primary protons and nuclei.  The parameters of muon production versus slant depth in the atmosphere (\refeq{eq:formula}) are given  
for two regions of muon energy, $\sim$TeV and ${\sim}\SI{100}{\GeV}$ in Tables~\ref{tab:params} and \ref{tab:params_ND}.  
The production profile can be folded with
atmospheric temperature profiles to obtain the multiplicity and size of muon bundles in air
showers for which the primary energy is determined by a surface array (\refsec{sec:surface-underground}). 
Future application of the parameterization of \reftab{tab:params} to seasonal variations of muon events in IceCube requires implementation of the formulas in a framework
that accounts separately for the contribution from each bin of muon energy at
each depth to be folded with an effective
area function that depends both on direction and energy.

The production profiles can also be used
as a weighting factor to calculate effective temperatures for analysis of seasonal variations
in underground detectors, including events with multiple muons (\refsec{sec:rates}).  Our focus in the latter application is to  investigate the altitude effect by which parent mesons decay higher in the summer causing multiple muon events to be more spread out at the detector.  We find that this effect makes an important contribution to the observed anti-correlation with temperature of multiple muon events for both the MINOS FD at TeV energies and for the shallow detectors at lower energy.  With the muons more spread out, the rate at small separations decreases.

For the MINOS
FD~\cite{Adamson:2015qua} the size of the TeV muon bundles is comparable to the detector area.  In this case we find that the altitude effect reverses the correlation coefficient for the innermost selection region, and that the anti-correlation becomes more prominent when the additional effects of multiple scattering in the overburden and bending in the geomagnetic field are accounted for.  
At the shallow detectors at Fermilab with lower energy muons, the multiple muon events have a characteristic size much larger than the area of the detector and only the geomagnetic effect is significant.  For the NOvA ND we find an anti-correlation for all multiple muon events but with an amplitude significantly smaller than observed.  Our approximate results indicate the need in both cases for more extensive analyses that account for the details of detector acceptance and for the additional effects of multiple scattering and charge separation in the geomagnetic field.

\noindent {\bf Acknowledgments}: 
We thank Spencer Klein for reading an initial draft of the paper and pointing out the importance of multiple scattering and bending in the geomagnetic field for the analysis of multiple muon events.  We thank the referee who also noted the relevance of multiple scattering.
We thank Dennis Soldin, Segev BenZvi and Carlos Arg\"uelles for helpful comments on this work. S.V. acknowledges the Fund for Scientific Research-Flanders (FWO) and the FWO Big Science programme. S.V. appreciates the hospitality of the Bartol Research Institute, where the idea for this paper emerged.

{\footnotesize
\bibliography{G-V}
}

\end{document}